\documentclass[aps,prb,twocolumn]{revtex4-2}
\usepackage{amsmath,amssymb}
\usepackage{graphicx}
\usepackage{mathtools}
\usepackage{float}
\usepackage{color}

\begin{document}

  \begin{abstract}
    A two dimensional classical ferromagnetic XY model with its bound vortex-antivortex dominated quasi long range ordered phase at low temperatures is a long standing as well as well studied problem of interest in the field of condensed matter. We conduct a detailed Monte Carlo study of such model in a square lattice with rather unexplored extensions where additional anisotropic exchange coupling and Dzyaloshinskii–Moriya interactions (DMI) together affect the Kosterlitz-Thouless (KT) transition in presence/ absence of symmetry breaking fields. Without DMI, the exchange term promotes collinear (ferromagnetic) order, whereas the DMI term induces spin cantings. By tuning anisotropy upto Ising limit, we document energy, specific-heat, magnetizations as well as helicity modulus and vortex densities for different temperatures and DMI strength.
    We also compute the 2nd moment of correlation lengths in order to probe the spatial correlation of the spins. Furthermore, the effect of U(1) symmetry breaking 4-fold and 8-fold symmetric $h_4$ and $h_8$ fields are explored which shows how the double-peaked specific heat profiles changes in presence of DMI. Overall, our findings append many important updates in the low temperature phases of a topological XY ferromagnet when additional DMI and isotropy-breaking exchange and/or field terms are considered thereby providing a few practical blueprints for suitably engineering topological spin systems.
  	
  \end{abstract}

\title{Interplay of Anisotropy, Dzyaloshinskii Moriya Interaction and Symmetry breaking Fields in a 2D XY Ferromagnet}
\author{Rajdip Banerjee}\email{rajdip.banerjee9696@gmail.com}
\affiliation{Department of Physics, Indian Institute of Engineering Science and Technology , Howrah, West Bengal-711103, India}
\author{Satyaki Kar}\email{satyaki.phys@gmail.com}
\affiliation{Department of Physics, A.K.P.C. Mahavidyalaya, Bengai, West Bengal-712611, India}
	  \maketitle
\date{\today}
	\section{Introduction}
	A two dimensional (2D) ferromagnetic XY model which is the primary model to exhibit Kosterlitz Thouless (KT) transitions\cite{kt,kt0}, is considered sacred for its association with a quasi-long range ordered phase at low temperatures (T). As the Mermin-Wagner's theorem\cite{auerbach} forbids the existence of long range order at any finite temperature in a 2D model with continuous symmetric short-range interactions, the low temperature KT phase with quasi-long range order (qLRO) featuring bound vortex-antivortex pairs as topological excitations, carry utmost importance in the field of condensed matter physics. Above the critical temperature $T_{KT}$ such excitations unbind to produce a paramagnetic {(PM)} disordered phase where the correlation length decays exponentially\cite{3}. This model is very much capable of explaining superfluid transition\cite{4} or the physics of 2D Josephson junction arrays\cite{5}. Even such transitions have recently been observed in ultracold Fermi gas \cite{6}. Certain liquid-crystal or magnetic films exhibit XY-like behavior and KT criticality \cite{7}. {One can categorize 2D magnetic systems as members of van der Waals magnets where atomically thin magnetic materials become versatile platforms for realizing low-dimensional Ising, XY or Heisenberg-like spin physics\cite{Park2025,Gong2019} that witness even exotic features like spin-charge correlation, spin-orbit torques or multiferroelectricity\cite{njp-zhang}. In particular, $CrCl_3$ has been reported as a nearly ideal easy-plane system exhibiting critical scaling characteristic of a 2D XY magnet and a finite-size BKT transition~\cite{Park2025,Song2022}. Likewise, $NiPS_3$ is described as an XXZ antiferromagnet (with weak FM coupling along $c$ direction) showing dominant XY-like behavior at the low temperatures\cite{Wildes2022}. These realizations provide strong motivation for studying the effects of anisotropy and symmetry-breaking fields in a planar XY ferromagnet.}

        Decades of Monte Carlo studies on ever larger square lattices have thus been devoted to pinning down the critical temperature and universal exponents of a 2D XY ferromagnet (XYFM), providing a benchmark for experiments and for related planar ferromagnets\cite{8}. 
        In real magnetic materials, however, spin-orbit coupling can introduce additional anisotropic exchange terms that alter the simple XY physics. A central example is the Dzyaloshinskii–Moriya interaction (DMI), which arises microscopically from superexchange in crystals lacking inversion symmetry. The DMI term is antisymmetric in the spins and thus favors a fixed sense of chirality between neighboring spins. This interaction was first introduced by Dzyaloshinskii to interpret the weak ferromagnetism of the antiferromagnetic materials\cite{9}.
        In the context of a 2D planar ferromagnet (FM), adding a uniform DMI term effectively imposes an intrinsic twist on the XY spins, favoring non-collinear configurations with a well-defined handedness. Thus DMI enriches the XY model with chiral and spatially modulated magnetic orders. Recent theoretical and numerical work has begun to map out the phase behavior of the 2D ferromagnetic XY model with a DMI. Liu $et.~ al.$ performed large-scale Monte Carlo simulations on square lattices using a hybrid update scheme and special boundary conditions\cite{landau}. They introduced a novel order parameter sensitive to the DMI-induced spiral and adopted fluctuating boundary conditions (FBC) to accommodate the incommensurate pitch of the spin modulation\cite{landau}. Building on this, Silva et al. carried out extensive Monte Carlo studies of the same model, employing a single-spin Metropolis plus Swendsen–Wang cluster hybrid algorithm and single-histogram techniques\cite{12} and obtained $T_{KT}$ in excellent agreement with the known analytic relation for the twisted XY model with an incommensurate phase.

        On the theoretical side, a renormalization–group analysis based on the Coulomb–gas duality has shown that the Dzyaloshinskii–Moriya interaction enters the vortex description as an effective uniform electric field capable of driving the system toward vortex unbinding for sufficiently strong coupling\cite{13}. The thermal and critical behavior of this type of XY model with asymmetric exchange and different types of randomness conditions has been an active subject of Classical Monte Carlo (MC) studies in recent years\cite{olivia1,olivia2,olivia3}. Classical Monte Carlo (MC) simulations provide a direct, nonperturbative way to probe the equilibrium thermodynamics of lattice spin models. For two-dimensional planar magnets they are particularly well suited to study the interplay of topological defects and collective order. In the pure 2D XY model MC has long been the tool of choice to demonstrate Berezinskii–Kosterlitz–Thouless (BKT) physics. A RG analysis described the topological phase behavior of {quantum version of} anisotropic XY model with DMI\cite{17}.

        Despite these advances, several open questions motivate further investigation. Away from the SO(2) symmetric isotropic point, what could be the effect of DMI on thermodynamic properties in the two dimensional anisotropic XY ferromagnet? How the critical behavior changes with changing DMI strength there? The exchange anisotropy attempts to break the vortex excitations and so question arises on what kind of other order it may lead to. Notice that an exchange anisotropy only makes the system anisotropic {in the spin space and continuous U(1) spin-rotational} symmetry is lost giving a lower $Z_2$ symmetry.
The preferred discrete spin orientations thus produce a long range order (LRO) instead of a qLRO at low temperatures though at a high enough temperature, free vortices can exist in an anisotropic XYFM. However, the transition temperature $T_{c}$ increases with the strength of anisotropy as the stability of the vortex excitations need to overcome larger thermal energies. How such behavior is modified in presence of DMI is certainly worth investigating. There can be different genre of anisotropy as well which also lower system symmetries. 
Real 2D magnetic systems of atomic monolayers or crystalline surfaces, that mimic approximately the XY physics, often experiences anisotropic crystal fields breaking continuous spin symmetry to discrete $Z_p$ symmetries\cite{sb3}. So their behavior in presence of a DMI is also worth exploring. Hence a numerical Monte Carlo based study investigating the interplay of DMI, exchange anisotropy and symmetry breaking fields is long due which we attempt to address in this work.

The paper is organized as follows. In Section II we define our model Hamiltonians and outline the simulation protocols. Then in Section III we show the numerical results for a medley of choices in DMI, anisotropy and fields. Finally in section IV, we {summarize} and discuss our results also briefing on possible future direction of our work.

\section{MODEL AND METHOD}
    To investigate the finite-temperature behavior of a 2D XYFM and its extensions, we perform classical Monte Carlo simulations on a two-dimensional square lattice governed by a Hamiltonian incorporating XY-type spin exchange, bulk  Dzyaloshinskii–Moriya interaction (DMI) as well as exchange anisotropy.
    We use the Metropolis algorithm, a widely adopted Markov Chain Monte Carlo method\cite{berg}, which proceeds by selecting random spin configurations and updating it probabilistically with temperature dependent Boltzmann acceptance probability of $P = \min\left[1, \exp\left(-\frac{d E}{k_B T}\right)\right]$
    where $dE,k_B$ and $T$ represent the energy difference between the updated and initial spin configurations, the Boltzmann constant and the temperature respectively. In our simulations we usually start from very high temperature $(\beta\sim0$) and gradually cool down to low temperatures. This updating scheme satisfies the detailed balance and ergodicity, ensuring convergence to the correct equilibrium ensemble. To improve statistical reliability, we perform $2\times10^5$ Monte Carlo steps/sweeps (MCS) at each temperature point, discarding the first $10^5$ steps for thermalization {(see appendix). 
Since the focus of the paper is on equilibrium thermodynamic properties rather than dynamical critical behavior, we consider Metropolis sampling with sufficiently long thermalization steps, like several MC studies on similar systems that used simple single spin Metropolis updates alone to produce meaningful results\cite{olivia1,olivia2,olivia3,Maccari2020}.} Periodic boundary conditions are imposed to suppress edge effects and mimic an infinite lattice. From the equilibrium configurations obtained from the MC simulations, one can compute ensemble-averaged thermodynamic quantities such as internal energy 
    $E = \langle H \rangle$ and specific heat 
    $C = \frac{d\langle E \rangle}{dT}$ and so on, which allow one to identify the location and nature of thermal phase transitions.

    Let us first consider a basic 2D isotropic ferromagnetic XY model\cite{18}whose Hamiltonian is given by,
    \begin{equation}\label{eq:iso_ham}
       H_{xy}=-\sum_{<i,j>}J{\vec{S}_i \cdot \vec{S}_j}
         =-J\sum_{<i,j>}\cos(\theta_i-\theta_j)
   \end{equation}     
      \\Each site (i,j) on a square lattice carries a planar (XY) spin represented by an angle $\theta_i$, and the Hamiltonian represent a ferromagnetic exchange between nearest neighbors, simplified as cosine functions of relative spin orientations among the nearest neighbor sites. In practice we sample spin configurations at various temperatures $T$ by randomly updating individual spins and accepting or rejecting moves according to the Boltzmann weight $\exp[-d E/(k_B T)]$.
     The specific heat per spin is computed from energy fluctuations using the fluctuation–dissipation theorem\cite{fluc-dis} as,
    \[
    C_V = \frac{{k_B}}{N} \beta^2 \left( \langle E^2 \rangle - \langle E \rangle^2 \right), \quad \beta = \frac{1}{k_B T}
    \]  where $N=L^2$, the number of sites in the square lattice. This can be evaluated following standard classical Monte Carlo simulations. 

    Exchange anisotropy can be introduced in such model to investigate whether and how the BKT transition survives with anisotropy. The corresponding Hamiltonian becomes
    \begin{equation}
   	H=-\sum_{<i,j>}(J_x{S_i^x  S_j^x}+J_y{S_i^y  S_j^y}) \label{eq:aniso_ham}
   \end{equation} 
   \\
   As per the standard protocol,  we can reduce this two-parameter ($J_x,J_y$) anisotropic model into an effective single parameter ($\Gamma$) model\cite{olivia1} expressed as,
    \begin{equation}
    	H_{a}=-J\sum_{<i,j>}[(1+\Gamma){S_i^x  S_j^x}+(1-\Gamma){S_i^y  S_j^y}], \label{eq:aniso2_ham}
    \end{equation}
    $J$ being a common scale factor.

    Next we introduce DMI in this anisotropic model. For that we consider a canonical transformation, as outlined by Liu et al.\cite{landau}, where the effect of DMI can be easily assimilated in the Hamiltonian of a 2D XY ferromagnet as
     \begin{align}
    	H_{ad}&=H_a-\vec{D}\cdot \sum_{<i,j>}{(\vec{S}_i\times\vec{S}_j)}\nonumber\\
    	=-&J[\sqrt{1+d^2}\sum_{<i,j>}\cos(\theta_i-\theta_j-\phi)-\Gamma\sum_{i,j}\cos(\theta_i+\theta_j)]
        \label{eq:dmi_ham}
     \end{align}	
     The DMI vector here is taken along the $z$ direction. Here $d=D/J$ denotes the strength of DMI and the angle $\phi=sin^{-1}(\frac{d}{\sqrt(1+d^2)})$ .
     Apart from specific heat, one can also estimate magnetizations of the system.
     As the spins on each site has two components, the average magnetization is calculated as
     \begin{equation}
       <m>=\sqrt(m_{x}^2+m_{y}^2)~~~~\rm{ with}~~~~\{m_x,m_y\}=\frac{1}{N}\sum_{i=1}^{N}\{S_{i}^x,S_{i}^y\}.
       \label{m-eq}
     \end{equation}
     We adopt periodic boundary conditions for the classical Monte Carlo simulation on L$\times$L square lattice with total sites, $N=L^2$. To avoid incommensurability due to boundary\cite{landau}, we consider system sizes in multiple of $8$ such as $L=8,16,24,32,40,48$ and consider commensurate DMI with $d=1/\sqrt{3},1$ and $\sqrt{3}$\cite{landau}.
     
      Now it has been found that  the $XY$ model exhibit interesting phase behaviors in presence of competing symmetry breaking fields (SBF) with strengths $h_4$ and $h_8$ (with 4-fold and 8-fold rotation symmetry respectively to break the SO(2) symmetry of the 2D isotropic $XY$ model)\cite{sb1,sb2,sb3}. Hence behavior of these fields in a 2D XY model in presence of DMI and exchange anisotropy, as represented by the Hamiltonian
      \begin{eqnarray}
        H_{adh}=H_{ad}-\sum_{j}h_j\sum_{i}\cos(j\theta_i).
      \end{eqnarray}
      can be worth exploring.
      Notice that a magnetic field, namely a $h_1$ SBF does not produce any nontrivial change in phase other than pushing for a collinear order, like in the Ising case. Hence here we rather report the nontrivial outcomes of introducing $h_n$ fields $(n>1)$ in our system and study the noteworthy changes in the phases and particularly, the $C_V$ or $m$ plots with temperatures.
      
     \section{RESULTS} 
     In this section we will show our simulation results step by step starting from the isotropic XY model.
     \subsection{Isotropic XY Ferromagnet} 
\begin{figure}[ht]
       \includegraphics[width=0.59\linewidth]{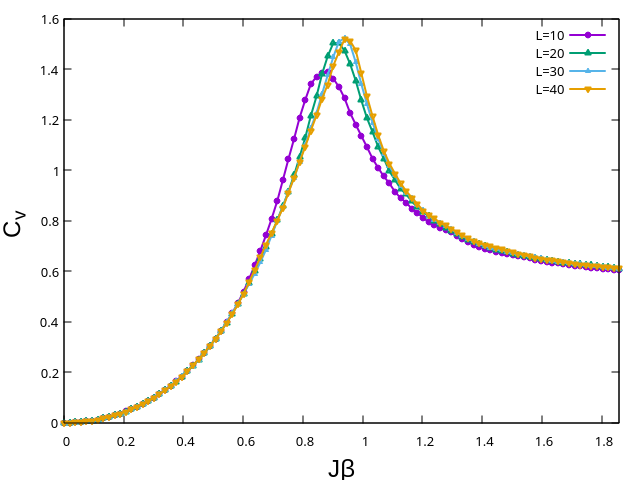}
               \put(-105,80){(a)}
\\
       \includegraphics[width=0.7\linewidth]{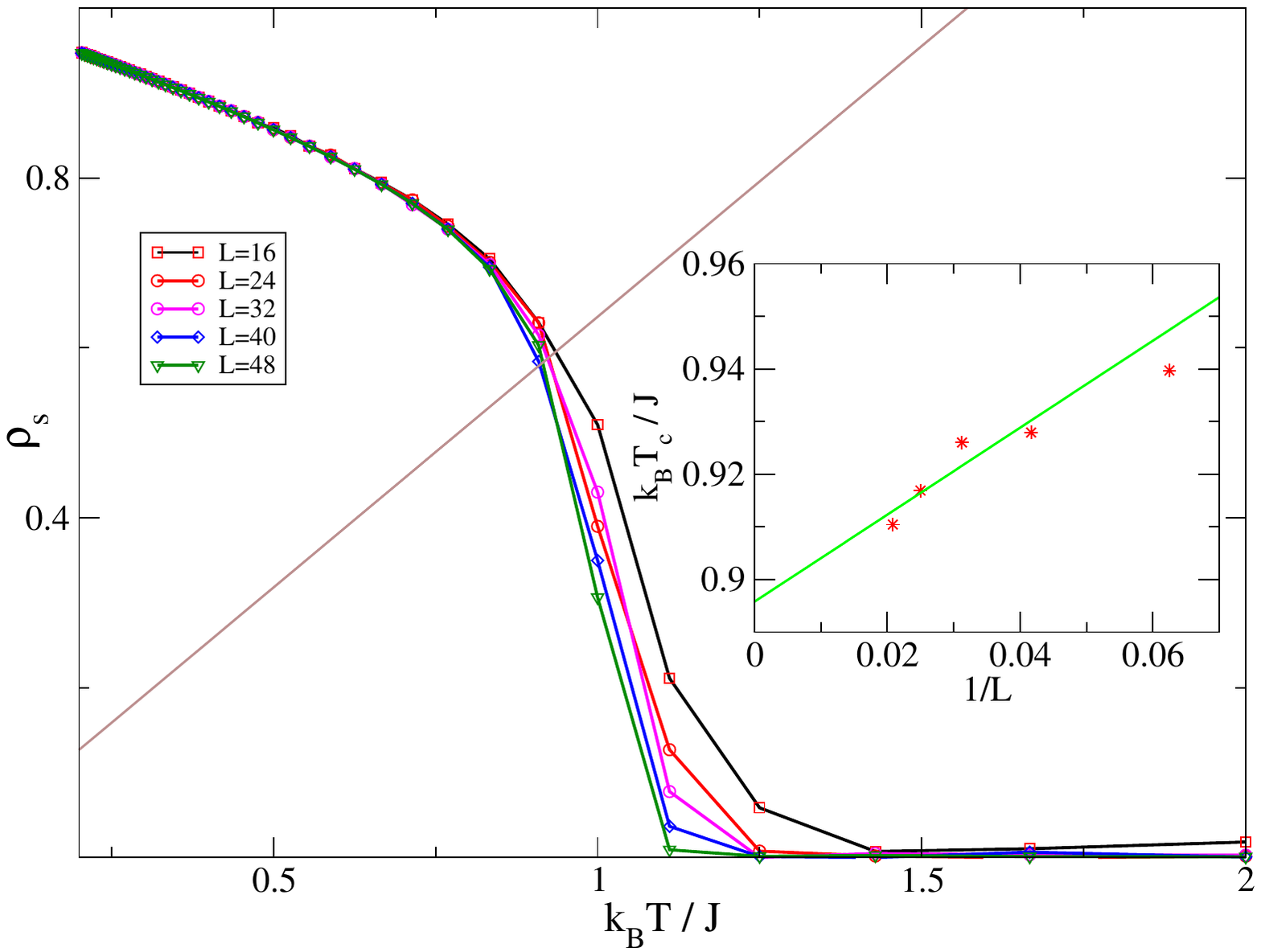}
  \put(-120,100){(b)}
    	\caption{(Color online)(a) Specific heat $C_V$ and (b) spin stiffness $\rho_S$ as a function of $J\beta$ and $k_BT/J$ respectively for different lattice sizes. The intersecting line in (b) corresponds to $\rho_S=\frac{2}{\pi}T$.}
    	\label{fig1}
    \end{figure}

    The thermodynamic behavior of a two-dimensional ferromagnetic XY model has been widely studied using classical Monte Carlo simulations. Nevertheless, for the sake of the continuity of discussion, we begin our calculations for such isotropic case where gradually cooling down the system from high temperature random spin configurations, we witness the transition from a paramagnetic phase to a phase with qLRO, where no long range order is observed though spin-spin correlation shows a slower power-law decay. In Fig.\ref{fig1} we capture the system behavior in specific heat $C_V$ in terms of inverse temperature $\beta$, expressed in units of $k_{B}/J$.
     \\As the temperature is lowered, the system undergoes a Berezinskii–Kosterlitz–Thouless (BKT) transition.     The rapid decay of $C_V$ on the low-temperature side reflects the essential singularity of the KT transition\cite{raghav}.
The U(1) symmetric Hamiltonian does not accompany any symmetry breaking in the transition but the correlation function undergoes a change in its functional form with the correlation length being infinity in the low temperature phase\cite{raghav}.
     In this low-temperature quasi-ordered phase, bound vortex–antivortex pairs are formed in the lattice which gets unbound at high temperatures due to thermal fluctuations. At low temperatures, spin-wave excitations also co-exist with the vortex-antivortex pairs\cite{PRB16-1217}, though this transition in 2D is mostly dominated by the unbinding of vortices though the spin stiffness is lost in the high temperature phase\cite{prb102-104505}. The low T spin-spin correlation function usually shows a power-law decay as $<S(r).S(0)>\sim r^{-\eta}$ with the temperature dependent exponent $\eta=\frac{k_BT}{2\pi J}$ that indicates a line of critical points for $0\le T\le T_{KT}$\cite{goldenfeld}. Notice that for a single vortex to be energetically favorable due to free energy gain, one can estimate the transition temperature to be $T_c=\frac{\pi J}{2k_B}$, at which the exponent $\eta$ takes its maximum value of $1/4$\cite{goldenfeld}. This, however, over-estimates the KT transition temperature as proliferation of vortices and their interactions are not considered in its naive derivation. The low-T phase with qLRO is generally characterized with condensation of topological defects appearing in the form of bound vortex-antivortex pairs in the system, contrary to the plasma of unbound vortices at high temperatures. Thus the KT transition is an entropy driven transition where a single vortex become unstable at low temperatures and can exist only by binding with antivortices\cite{goldenfeld,simons}.
     
   Notice that topological phases occur in the XY model at all finite temperatures due to formation of vortices/antivortices and thus the phase transition is not a topological phase transition in the conventional sense. Rather it only involves binding/unbinding of the topological defects.  In this respect, one can compute the spin stiffness $\rho_S$ (also called the helicity modulus) that measures how much free energy it costs to twist the spin angles slowly across the system. In other words, it tells you how rigid the phase field is against a long-wavelength distortion.  {\it The transition temperature ($T_{KT}$) is characterized by a finite jump in  $\rho_S$}\cite{bravo} (see Fig.\ref{Fig.2}(c))  that usually occur at $T=T_{KT}\simeq0.893J/k_B$ while a broad peak in the specific heat $C_V$ (as shown in Fig.\ref{fig1},\ref{Fig.2}(a)) appears due to proliferation of free vortices at a higher temperature. Usually, $\rho_S$ at the transition becomes $\frac{2}{\pi}T_{KT}$\cite{vojta} and with finite size analysis we estimate the transition to be at $T_{KT}\approx 0.895J/k_B$ in the thermodynamic limit. MC calculations are also carried out for several lattice sizes, including $L=10,20,30,$ and 40 to capture $C_V$ peaks which appear at near about same temperature for all the system sizes considered with periodic boundary conditions(PBC). All the curves approach $C_V\sim0.5$ at low temperatures, as expected from the equipartition theorem.   The results can be verified from earlier MC simulation results\cite{19}. 
     
    \subsection{Anisotropic XY Ferromagnet}
    Next we consider exchange anisotropy in the model and the variation of transition temperature because of that. In particular, we probe the evolution of $C_V$ peak temperature as anisotropy parameter $\Gamma$ is varied through.
    \begin{figure}[ht]
      \includegraphics[width=\linewidth,keepaspectratio]{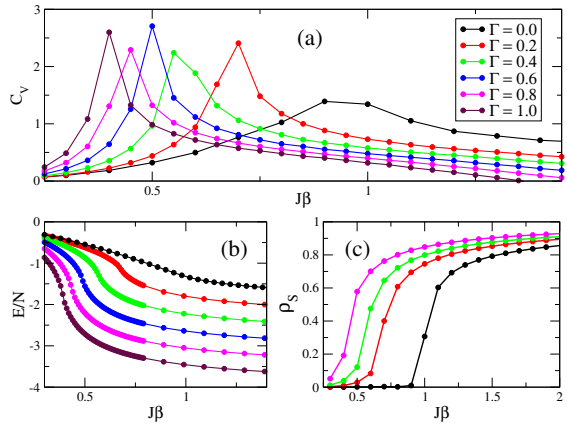}
      \caption{(Color online)(a) Specific heat $C_V$, (b) Energy per spin $<E>/N$ and (c) spin stiffness $\rho_S$ vs inverse temperature $J\beta$ for various $\Gamma$'s.}
      \label{Fig.2}
    \end{figure}     
         
    Fig.~\ref{Fig.2} shows the results of Monte Carlo simulations of the anisotropic two-dimensional XY ferromagnet with $L=48$ to illustrate how the strength of exchange coupling anisotropy modulates the thermal and topological excitations. Notice that the broad peak for $C_V$ gradually becomes sharp with an increase in anisotropy
    , indicating higher-temperature Ising like transitions for larger $\Gamma$ values. At the Ising limit of $\Gamma=1$ we get a transition temperature to be at $\beta\sim0.4/J$ compared to the theoretical estimate by Onsager's solution\cite{onsager} of $\beta_c=0.45/J$ for a 2D Ising model in a square lattice. The average energy per spin $\langle E \rangle/N$ decreases monotonically with increasing $\beta$, reflecting the progressive development of spin alignment due to suppression of thermal fluctuations. At high temperature, thermal disorder dominates and the energies become nearly independent of the exchange coupling and in particular, the anisotropy parameter $\Gamma$.
    
    Contrarily upon cooling, anisotropic effects become discernible as a smaller $J_y$ (or, larger $\Gamma$) allow spins to align more along the stronger bond directions $\hat x$ thereby reducing the energy. Each energy curve exhibits a concave profile that flattens at large $J\beta$, tending towards its ground-state value.
    Notice that, with $\Gamma\ne0$,the system carries a finite magnetization (which is not just a finite-size effect) at the low temperatures with $\langle m\rangle\ne0$ and thus a true long range order appears there.
    Interestingly as the anisotropy strength becomes nonzero and the system goes from a BKT phase to a LRO Phase, the spin stiffness also remains nonzero only to vanish in the high T paramagnetic phase (see Fig.\ref{Fig.2}(c)).
    In the isotropic case, spins have no preferred orientations.
   But that freedom is reduced, when exchange anisotropy is introduced. And due to less number of available states at low energy, the entropy is reduced as well. As a result, it takes a higher temperature to destroy the ordered or quasi-ordered state. Hence we find an increase in $T_c$ with increase in anisotropy.

    At the isotropic point, the specific heat $C_V$ displays a broad, muted hump around $J\beta\approx0.85$. 
    Increasing anisotropy shifts the specific-heat peak to smaller $\beta$ values while amplifying its sharpness and height.     Shifting of $C_V$ peaks towards larger temperature with an increase in anisotropy marks a shift in universality class of transition from XY to Ising type.


   \subsection{Anisotropic XY Ferromagnet with Dzyaloshinskii Moriya Interaction}
   Introducing a Dzyaloshinskii–Moriya (DM) term to the anisotropic two-dimensional XY model further alters the spin texture as well as the related thermodynamic properties.
   The effect of DM interaction in an isotropic 2D XY model has been studied following analytic simplifications and classical MC simulations\cite{landau}. We would like to follow those steps in presence of exchange anisotropy.
   
     We consider the Dzyaloshinskii-moriya vector ${\bf D}$ to be perpendicular to the $xy$ plane. Firstly in the isotropic limit, with the anisotropy parameter $\Gamma = 0$, we find the variations of transition temperature as shown in the Fig.\ref{fig:iso_dmi}.
     \begin{figure}[ht]
       \includegraphics[width=\linewidth]{Cv-with-d.eps}
     	  \caption{(Color online)Specific heat $C_V$ as a function of inverse temperature $J\beta$ for DMI strengths {$d=0,\frac{1}{\sqrt{3}},1$ and $\sqrt{3}$} in the 2D isotropic XY ferromagnet. Inset shows the Phase Diagram {featuring low-$T$ QLRO and high-$T$ PM} phase.}
     	  \label{fig:iso_dmi}
     \end{figure}
     From there, it is evident that increasing DMI strength increases the critical temperature and the new spin-canted qLRO lasts longer against thermal fluctuations (see Fig.\ref{fig:iso_dmi} inset).
    
     The $d$ vs $T_{c}$ phase diagram portrays how the $KT$ transition temperature gets modified in presence of DMI. Usually DMI leads to spin canting where the nearest neighbor spins no longer remain parallel to each other. Rather they prefer an angular deviation of $\phi$, particularly in the isotropic limit (see Eq.\ref{eq:dmi_ham}).
As we only deal with PBC, rather than a fluctuating boundary condition\cite{landau}, the lattice sizes should be restricted in order to avoid any undue boundary effect coming from DMI and its resulting spin cantings. In fact we only want to deal with commensurate spin cantings. For a square lattice with a length of $L$, the modulated spin configuration should be such that, it satisfies $L/a=2n\pi/\phi$, $n$ being an integer. And as $\phi=tan^{-1}(d)$, not every $d$ values can give commensurate spin configurations. We find that $L=48$ turns out to be a commensurate lattice size for $d=1/\sqrt{3}$, 1 or $\sqrt{3}$.
      \begin{figure}[t]
        \begin{center}
        \begin{picture}(100,100)
          \put(-70,0){
     \includegraphics[width=.49\linewidth,keepaspectratio]{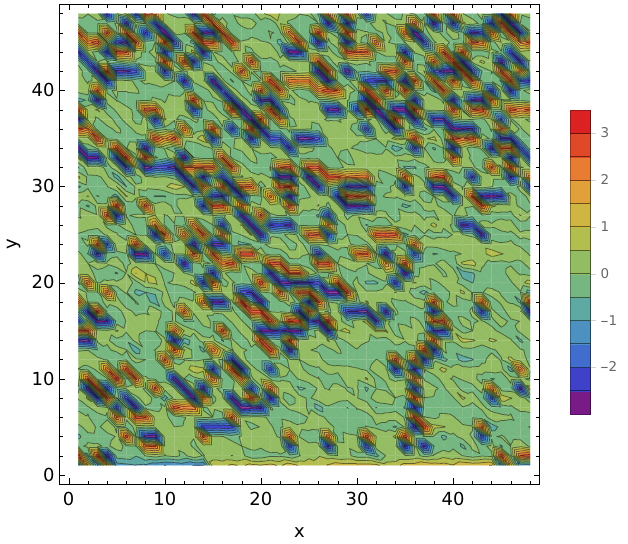}
     \includegraphics[width=.49\linewidth,keepaspectratio]{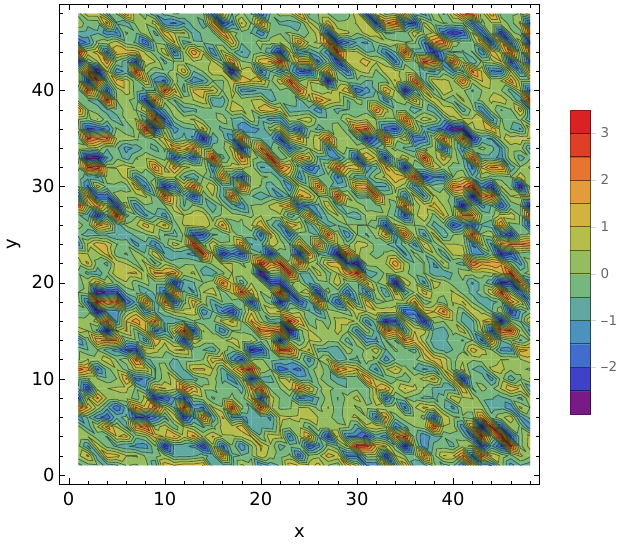}}
          \put(-70,-110){
\includegraphics[width=.493\linewidth,keepaspectratio]{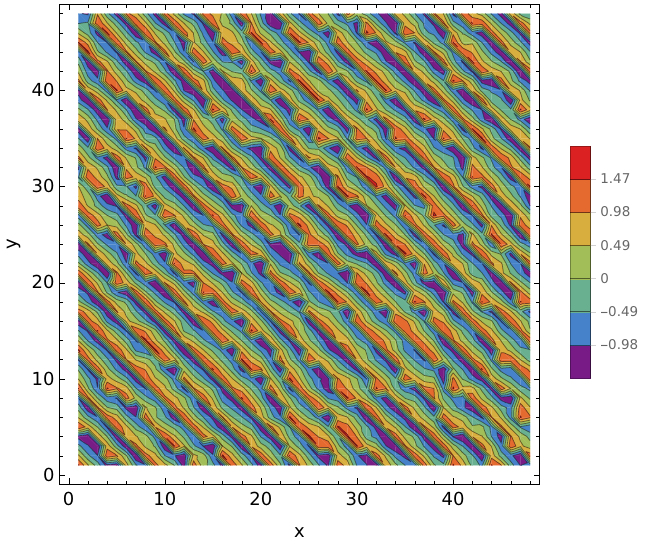}
     \includegraphics[width=.495\linewidth,keepaspectratio]{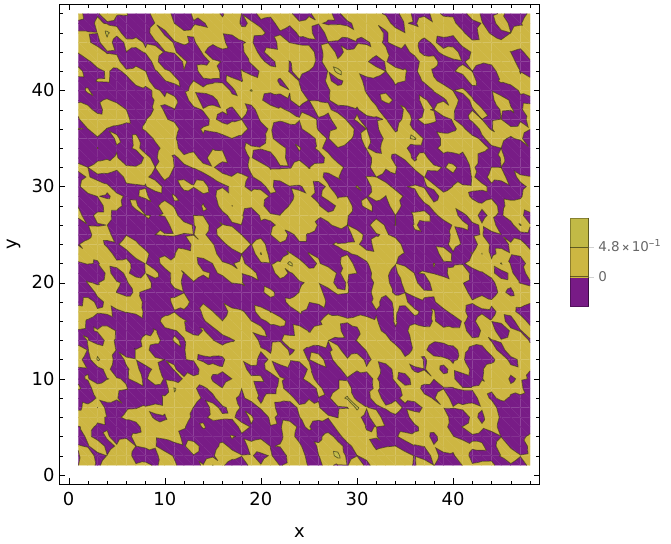}}
     \put(-75,95){(a)}
     \put(50,95){(b)}
     \put(-75,-20){(c)}
    \put(50,-20){(d)}
          \end{picture}
        \end{center}
 \vskip 1.4 in    
  \caption{(Color online)Variations of typical spin configurations in terms of $\partial_x\theta$ for $(d,~\Gamma)~=$ (0,0) at (a) $T=0.2$, (b) $5.6$ (in units of $J/k_B$) respectively while typical spin configurations $(\theta)$ at  $T=0.2$ for $(d,~\Gamma)~=$ (c) (1,0) and (d) (0,0.4) respectively.}
     	\label{fig:iso_dmi_spin}	
      \end{figure}
      The 2D isotropic XY model show a FM LRO at $T=0$. However, as per Mermin-Wagner theorem $<m>=0$ at any finite temperature in a 2D XY model in thermodynamic limit. Numerically for a small $T=0.2J/k_B$, bound vortex-antivortex pairs are observed in different pockets in the 2D lattice. It's better discernible if we plot the spin deviations or $eg.$, $\partial_x\theta$ within the lattice (see Fig.\ref{fig:iso_dmi_spin}(a)). It shows many local clusters of vortex-antivortex (with positive and negative $\partial_x\theta$ values respectively) pairs. {In fact, a small-magnitude value in the Fig.\ref{fig:iso_dmi_spin} color scales indicates a FM-like region while a comparatively large positive/negative
value indicates counter-clockwise/clockwise rotation of spin vectors relative to its x neighbor on the left.} At a high temperature, vortices and antivortices are more scattered through the lattice and do not appear as local pairs (see Fig.\ref{fig:iso_dmi_spin}(b)).     { Fig.\ref{fig:iso_dmi_spin}(c) gives typical spin patterns in a low-$T$ quasi-ordered phase for $d=1$.}
      Finite DMI strength gives a diagonal arrangement of spins (along $\hat{x}-\hat{y}$ direction) at low temperatures because in a minimum energy configuration, the positive $x$ and $y$ neighbor of each lattice point should have same spin cantings (as per Eq.\ref{eq:dmi_ham}) in the ground state. But this gets disrupted with temperature when topological excitations of vortex-antivortex pairs appear in the lattice. {Similarly, a low-$T$ anisotropic system prefers collinear alignments; more precisely, FM-like spin orientations along the direction of stronger exchange interactions (see Fig.\ref{fig:iso_dmi_spin}(d)).}

\begin{figure}[ht]
       \includegraphics[width=\linewidth,keepaspectratio]{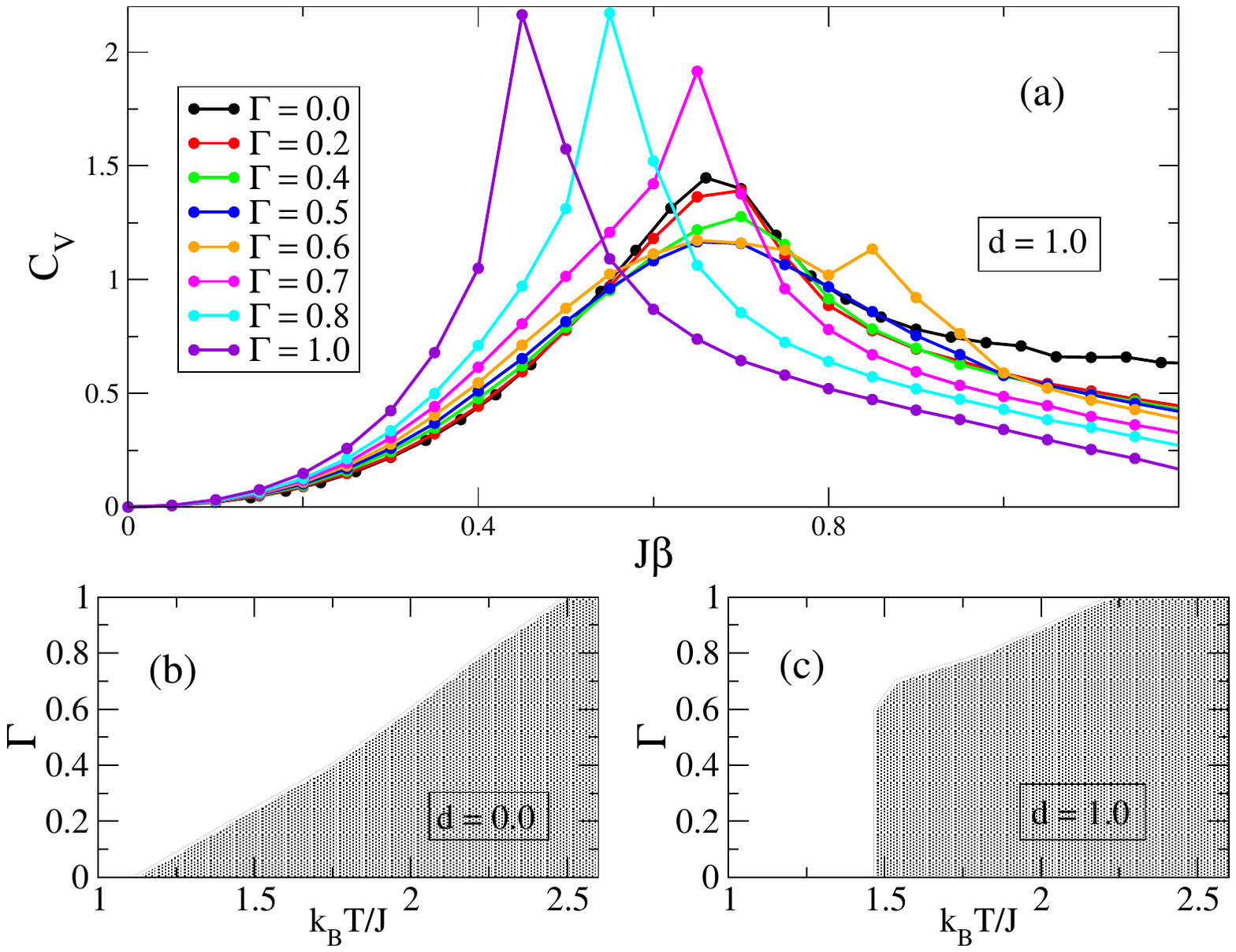}
         \caption{(Color online)(a) Specific heat $C_V$ as a function of inverse temperature $J\beta$ for $d=1.0$ and different $\Gamma$. (b), (c) Phase diagrams based on $C_V$ peak positions for $d = 0$ and 1 respectively. The shaded region indicate the high-T paramagnetic phases.}
     	 \label{fig:ani_dmi_spec}
     \end{figure}

      Away from isotropic limit with $\Gamma\ne0$, the system tends to align spins ferromagnetically along the direction of stronger exchange interaction and a magnetization ${\bf m}$ develops even at low temperatures. This however decreases in presence of DMI that tilt the spins away from the ${\bf m}$ direction. In an anisotropic XYFM, change in transition temperature (more precisely, the $C_V$-peak temperature) is rather low in presence of DMI. Fig.~\ref{fig:ani_dmi_spec} shows the variation of $C_V$ with anisotropy ($\Gamma$) as well as the corresponding phase diagram for $d=1.0$. The evolution of $C_V$ profiles are similar to that for $d=0$ case (compare with Fig.\ref{Fig.2}) though transitions occur at a bit higher temperatures.{Fig. \ref{fig:ani_dmi_spec}(a) shows that for d = 1.0 and smaller $\Gamma$ values, the peak remains broad and rounded, which is consistent with KT-like behavior dominated by continuous-spin fluctuations. It also manifests a reluctance in the change in $T_c$ upto $\Gamma\sim~0.6$. With further increase in $\Gamma$, the $C_V$ peak becomes progressively sharper indicating that the exchange anisotropy increasingly suppresses the continuous U(1)-like symmetry and drives the system toward an Ising-like regime.}
    \begin{figure}[ht]
     			\includegraphics[width=\linewidth,keepaspectratio]{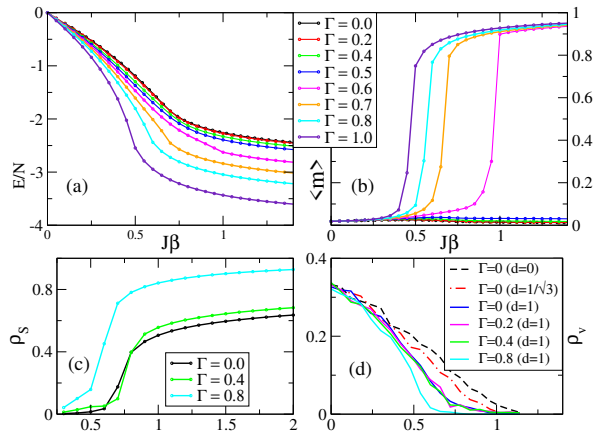}
     	\caption{(Color online)(a)Average Energy per spin, $\langle E \rangle/N$ and (b)Average magnetization $\langle m \rangle$ and (c) spin stiffness $\rho_S$ as a function of inverse temperature $\beta$ at different anisotropy strengths for DMI strength $d=1.0$. (d) shows the vortex density $\rho_v$.}
     	\label{aniso_dmi}
     \end{figure}
             
        
       In Fig.\ref{aniso_dmi}(a), we have shown energy per spin plotted against inverse temperature $\beta$.
        At small $\beta$, all energy curves tend to coincide close to zero because  thermal fluctuations randomizes spin orientations and washes out any interaction dependence. On cooling the system (increasing $\beta$) gradually sees the interaction dominance and drives the system to lower energies as expected \cite{19}. Furthermore, anisotropy helps building the collinear FM arrangement causing further decrease in the energy. The DMI term, contrarily, favors a chiral alignment. So with both DMI and the anisotropy present, the system witness a competition between these two effects and {university class of the} transition is affected accordingly.
     
      \subsubsection{Magnetization}

     In Fig.\ref{aniso_dmi}(b) we show the variation of magnetization order parameter with inverse temperature for different anisotropy strength with DMI.
   We calculate magnetization as ${\bf m}=\sqrt{m_x^2+m_y^2}$ where $N$ is the total number of lattice sites and $(m_x,m_y)=\frac{1}{N}\sum_{i}(\cos\theta_i,~\sin\theta_i)$, $i$ being the $i^{th}$ lattice site. Here we have considered, $N = 48\times 48$. Notice that at $d=1$ that we consider, it takes a large anisotropy with $\Gamma\sim 0.5$ to get finite magnetization at low temperatures. Thus, in contrast to the $d=0$ case, LRO doesn't appear immediately with anisotropy for $d=1$.  The transition of $\Gamma=0$ gets modified with $\Gamma\ne0$ as anisotropy pushes for ising-like configurations in the system providing nonzero magnetization at low temperatures (though vortices still persists in presence of weak anisotropy). {In fact, numerical results also indicate finite vortex densities at high temperatures (see Fig.\ref{aniso_dmi}(c)). Thus vortices do not readily vanish at the advent of anisotropy. }
 $\Gamma$ systematically shifts the magnetization onset to smaller $\beta$ (higher $T$) and sharpens the crossover, indicating that anisotropy suppresses low-energy spin-wave fluctuations and stabilizes order due to breaking of the spin rotational SO(2) symmetry of the isotropic ferromagnet. All curves saturate close to unity ($\approx0.90 -0.95$) at low temperature, which we attribute to DMI-induced canting and residual chiral inhomogeneity that compete with exchange and anisotropy as exchange and anisotropy favor coherent alignment while DMI promotes twisted configuration.

   \subsubsection{Spin Stiffness and Vortex Density}
   For a XY model with DMI, Ref.\cite{landau} devised a magnetization-like order parameter where an effective magnetization is calculated taking into consideration the spin cantings due to DMI. But with exchange anisotropy in addition, such definition no more appropriately describe the order (due to the additional $2^{nd}$ term in the Eq.\ref{eq:dmi_ham}).
   As we mentioned earlier, calculating spin stiffness $\rho_S$ can become useful in these cases that can indicate the transitions even in presence of DMI, as shown in Fig.\ref{aniso_dmi}(d) for $d=1$.
In fact spin stiffness or helicity modulus is a natural measure of phase rigidity because it is defined through the second derivative of the free energy with respect to an imposed twist, and therefore quantifies the energy cost of a global deformation of the spins (\cite{vojta}). This quantity is especially significant in two dimensions, where the low-temperature phase is not characterized by conventional long-range magnetic order but by quasi-long-range order with finite stiffness; physically, a nonzero helicity modulus indicates that the system still resists twisting, while its rapid reduction with increasing temperature reflects the growing influence of vortex excitation and its fluctuations and eventual unbinding of vortex-antivortex pairs (\cite{sun}). Consequently, the temperature dependence of the helicity modulus is one of the most useful diagnostics of the BKT transition in the XY model, and in the thermodynamic limit it obeys the Nelson–Kosterlitz universal relation 
$\rho_S(T_{KT})=2T_{KT}/\pi$, so that the intersection of $\rho_S(T)$ with $2T/\pi$, provides an estimate of the transition temperature(\cite{nelson}).

In addition to spin stiffness, one can also calculate vortex densities ($\rho_v$) to identify the KT or KT-like transitions in such systems. This gives null vortex density $\rho_S$ in the low-T phase as only the bound vortex-antivortex pairs, rather than single vortices are formed in the system in isotropic and weakly anisotropic limit.
   Fig.\ref{aniso_dmi}(d) demonstrates such $\rho_v$ variations for such systems with and without DMI.

   \subsubsection{Second-Moment Correlation Length}
	
	To quantify the spatial extent of spin correlations in the system, we use the second-moment correlation length\cite{sb3,viet,komura}. For a finite two-dimensional lattice of linear size $L$, {we consider the second-moment correlation length to be
	\begin{equation}
		\xi^{(2)} = \frac{1}{2\sin(\pi/L)}
		\left(\frac{\langle m_1({\bf 0})^2\rangle}{\langle m_1({\bf k_m})^2\rangle} - 1\right)^{1/2},
	\end{equation}
		where $ m_1({\bf k})^2=\sum_{\mu=x,y}|\frac{1}{N}\sum_{i=1}^NS_i^\mu {\rm exp}(i{\bf k.r_i})|^2$ denotes the ${\bf k}$-dependent magnetization and ${\bf k_m}=(2\pi/L,0)$ is the smallest nonzero wave vector in the periodic lattice\cite{19,sb3}}. This quantity provides a compact measure of the characteristic length scale over which spins remain correlated. Unlike the full distance-dependent correlation function, which must be examined for different distances, the second-moment correlation length summarizes the overall correlation behavior via a single number\cite{viet}. This makes it especially useful in finite-size numerical studies, where direct analysis of the correlation function may be affected by noise and boundary effects.
\begin{figure}[t]
     			\includegraphics[width=\linewidth,keepaspectratio]{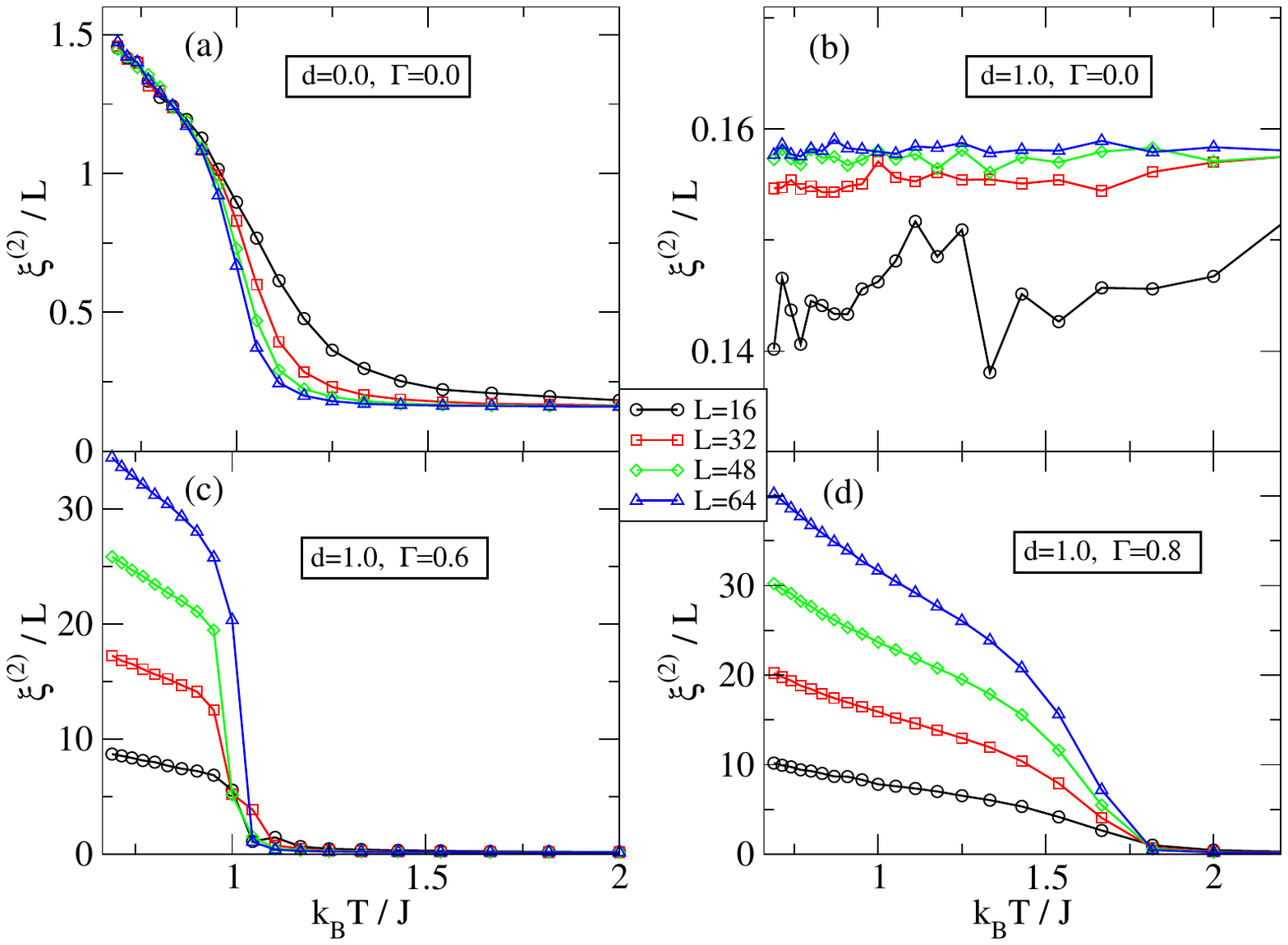}
      	\caption{(Color online) Variation of 2$^{nd}$ moment Correlation length with temperature for different DMI and anisotropy strengths.}
     	\label{cor}
     \end{figure}

The second-moment correlation length is particularly significant in the two-dimensional XY model because the system does not exhibit conventional long-range magnetic order in the usual sense. Instead, the thermodynamic behavior is governed by the evolution of correlations over different length scales. As temperature changes, the correlation length provides a direct indication of how far spin coherence extends in the system. Near the transition region, the growth or reduction of $\xi^{(2)}$ reflects the changing influence of thermal fluctuations and topological excitations. For this reason, it is a very useful quantity for identifying finite-size signatures of the Berezinskii-Kosterlitz-Thouless type behavior.
	
	In the present work, the second-moment correlation length has been used to compare the thermal behavior of several related models: the two-dimensional isotropic XY model, the XY model with Dzyaloshinskii-Moriya interaction (DMI) for $d=1$, and the DMI model with exchange anisotropy for $d=1$ and $\Gamma=0.2$, $0.6$, and $0.8$. For each case, simulations were performed on lattices of size $L=16$, $32$, $48$, and $64$. This systematic comparison allows us to examine how the inclusion of DMI and anisotropy modifies the correlation length and how the finite-size behavior evolves relative to that of the isotropic XY model.
        In Fig.\ref{cor}(a) the usual behavior of 2D XY model is shown where the correlation ratio $\xi^{(2)}/L$ remains finite and same for all the system sizes investigated at low temperatures but differs while dropping at a certain temperature. Such difference, for $d=0.0$, near transition region ($T\sim0.9-1.1$) is due to the finite size effect. At the high temperature, correlations become short-ranged. Interesting behavior is observed with the inclusion of DMI ($d=1.0$) (Fig.\ref{cor}(b)), as the correlation ratio remains close to zero at all the temperatures for all system sizes. This behavior reflects the lack of ferromagnetic correlation in the system of XY model with DMI at all temperatures, as DMI prefers a chiral configuration. In the system of XY+DMI, when we introduce anisotropy and increase its strength slowly(Fig.\ref{cor}(c)(d)), the effect of DMI is suppressed and correlation length grows again to a higher value at low temperatures. But then it drops again close to zero at a certain higher temperature marking the transition from low temperature ordered to a high temperature disordered state. {Thus one can witness and distinguish KT-like and Ising-like transitions from different features in $\xi^{(2)}/L$ (compare Fig.\ref{cor}(a) and Fig.\ref{cor}c,d).}

   \subsection{XYFM with symmetry breaking fields}

   { In the presence of symmetry-breaking fields, the thermal behavior of the 2D XY ferromagnet gets modified substantially.}
   An in-plane magnetic field (or, a 1-fold symmetry breaking field $h_1$) along $x$ direction suppresses the $KT$ transition in an isotropic $XY$ model {favoring collinear spin alignments turning the sharp vortex-driven transition into a crossover\cite{Gouvea1990} that occurs at a higher temperature} as free vortices become energetically costly and less likely to appear. Moreover, anisotropy enhances the field sensitivity driving the system faster towards transition.
   \begin{figure}[ht]
       \includegraphics[width=\linewidth,keepaspectratio]{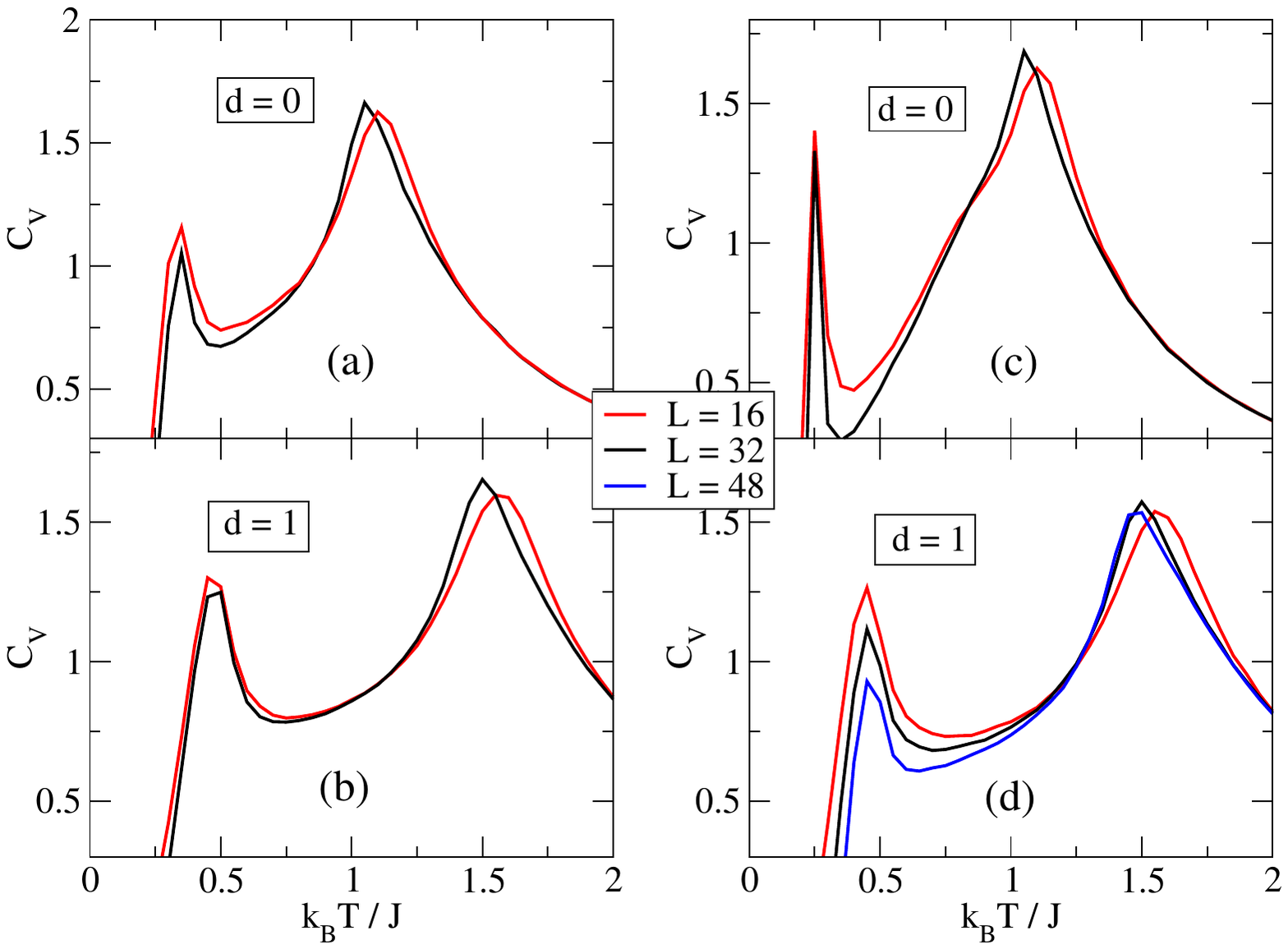}
         \caption{(Color online) $C_V$ as a function of $J\beta$ in presence of fields (a,b) $h_4=0,h_8=1$ and (c,d) $h_4=0.4,h_8=-0.6$ for (a,c) $d=0$ and (b,d) $d=1$ and different system sizes.}
     	 \label{sym-brk}
     \end{figure} 

   In presence of field $h_2$ or $h_3$ instead, the nature of transitions changes to Ising or 2-state Potts universality class respectively\cite{sb3}. { Thus, once discrete anisotropy is introduced, the low-temperature phase is no longer governed purely by the $U(1)$ symmetric KT physics, but by the competition between field-induced ordering and vortex-related excitations.}

Interestingly, with a $h_4$ field, specific heat features cusp-like peak (and no rounded peak typical of KT transitions) whereas  a $h_8$ field can produce double $C_V$ peaks. 
Particularly a $XYh_4h_8$ model\cite{sb3}, in both limits of compatible and competing spin configurations, deserves attention for it features double peaks in quantities like specific heat. A $h_4$ field giving a single cusp-like peak indicates a crossover to Ising-like phases. Contrarily for a $h_8$ field, {the $Z_8$ anisotropy is sufficiently weak which makes the system behave similarly to the $q=8$ clock model that supports an intermediate quasi-long-range ordered KT phase\cite{Tuan2022}. It gives two $C_V$ peaks (see Fig.\ref{sym-brk}(a)) that are consistent with two KT-related thermal scales, namely a low-temperature FM to KT crossover and a higher-temperature KT to paramagnetic transition\cite{sb3}.}

These two fields can prefer compatible or competing spin configurations depending on whether $h_4h_8>0$ or $h_4h_8<0$ respectively. In the compatible regime with both $h_4,~h_8$ positive, one sees a single high temperature continuous transition and/or a low-T crossover in addition\cite{sb3}. But in the competing regime of $h_4>0,h_8<0$, a low-T sharp peak {appears that  becomes more pronounced with system size} and a size-independent rounded KT-like peak at a higher temperature can be observed\cite{sb3}. Fig.\ref{sym-brk}(a) represents the temperature dependance of the specific heat for system size $L = 16, 32$
for $h_4=0.0$ and $h_8 = 1.0$ and $d = 0.0$. Here we have
used a total of $4\times 10^5$ Monte Carlo steps from which we neglected first $2\times 10^5$
steps
for thermalization and rest for measurements. Numerical results show two peaks, one at low temperature and the other one at comparatively
higher temperature, the peak heights being size independent. As mentioned before, the first peak associated
with low temperature indicates a transition from ferromagnetic to KT whereas the higher temperature peak corresponds to a transition from KT to paramagnetic
phase. The results are similar to that shown by Truong et al.\cite{sb3} where they have
implemented cluster updating to avoid critical slowing down near transition temperatures.

Interestingly, if we implement DMI into the system (see Fig.\ref{sym-brk}(b)), the transition temperature for both
the transitions get shifted to higher temperatures, the shift being larger for the high-T transition. {When DMI is included, the system no longer favors purely collinear alignment; instead, it prefers a twisted spin configuration with a fixed chirality. Firstly in the presence of symmetry breaking $h_8$ field alone, this chiral tendency competes with the discrete symmetry-breaking term, which try to lock the spins into specific angular directions. As a result, the thermodynamic anomalies in $C_V$ are shifted to higher temperatures, because a larger thermal energy is required to overcome the DMI-induced twisting. The higher-temperature peak is more strongly affected since it is associated with the loss of the topological KT-like correlations (in presence of chiral DMI), while the lower-temperature feature reflects the competition between discrete ordering and chiral canting. Thus, DMI reshapes the crossover features by stabilizing noncollinear spin textures.}

The isotropic
XY model in presence of $h_4 = 0.4$ and $h_8= -0.6$ (and without any DMI) that sees the two fields competing for their preferred spin configurations, also shows 2 peaks in $C_v$ representing a low temperature second order phase
transition that gets sharper with increasing system size and a high temperature KT
phase transition\cite{sb3}. The average magnetization remains nonzero till some finite temperature. Introducing DMI with this type of crystal field choice, however, results in flatter low-T transition, also dependent of system size (see Fig.\ref{sym-brk}(c,d)).
{This strongly suggests that the DMI competes with the discrete
anisotropy and destabilizes the low-temperature Ising-like phase, turning the sharp transition into a much weaker finite-size crossover in the thermodynamic limit. However, the high-temperature broad peak remains visible near the same temper
ature range and retains a KT-like character, indicating that the topological
transition is more robust against DMI than the low-temperature symmetry-
breaking transition.}
One can also find the average magnetization, in this case, to remain close to zero.
   
  \section{SUMMARY AND DISCUSSIONS}
   The Monte Carlo results presented in this work provide a clear and consistent picture of how directional exchange anisotropy, Dzyaloshinskii–Moriya coupling and symmetry-breaking fields reshape the finite-temperature behavior of planar ferromagnets. In the absence of DMI, the XYFM reproduces standard Berezinskii–Kosterlitz–Thouless (BKT) phenomenology and yields a transition temperature close to the well-known benchmark for the classical XY model ($\beta_{KT} \approx 1.12$), confirming the validity of our implementation and thermalization protocol\cite{19}. The anisotropy parameter ($J_y/J_x$ or $\Gamma$) systematically shifts pseudo-critical signatures. Stronger anisotropy makes the system softer towards one direction stabilizing a magnetic order. It drives the specific-heat peak to appear at higher temperature and become sharper, consistent with the fact that anisotropy breaks the continuous U(1) symmetry and favoring a two-fold classical degeneracy making the system an assembly of quasi-1D discrete spin chains, altering the balance between spin-wave and topological excitations.
   The interplay of exchange anisotropy and Dzyaloshinskii–Moriya coupling qualitatively modifies the thermal and topological signatures of two-dimensional planar ferromagnets. Anisotropy sharpens the crossover toward ordered behavior by partially lifting the degeneracy, while DMI stabilizes chiral, twisted spin textures and shifts pseudo-critical temperatures upward, an effect that have been reported in complementary Monte Carlo studies of XY+DMI models and captured by RG analyses of vortex gases in an effective background field\cite{21}. The combined numerical evidences, as reported here, support the conclusion that directional and chiral couplings are effective tuning knobs for pseudo-criticality in 2D magnets, with direct relevance for engineered ultra-thin magnetic films where DMI and anisotropy can be controlled experimentally\cite{22}. Furthermore, DMI also effectively tune the orders and transitions in presence of symmetry breaking fields that introduces spingaps\cite{fouet} in the system.  {One can also consider dipolar interactions\cite{dipole} that plays an important role in determining magnetic configurations in real materials and accordingly enhance the scope of the present study. However, as it requires introduction of long-range nonlocal terms which substantially increases the numerical complexity\cite{long-range}, we have kept aside such exploration for now and left it for future extension of the model.} We also plan to adopt a quantum version of such 2D XYFM based detailed analysis and extend that first for 3D Heisenberg spins so that behavior of skyrmions nucleated in such systems can be properly investigated to better understand the exotic phenomena like skyrmion Hall effects\cite{skyr}.

   \begin{appendix}
 {    \section{Monte Carlo Thermalization}
        In a typical MC simulation, one first wait for the system to thermalize corresponding to the given $\beta$ value and it is after that equilibrium properties are enumerated as ensemble averages. In the present work we discard initial $10^5$ MCS data for thermalization to take place and then measurements are made following the usual binning procedure. Fig.\ref{therm} demonstrates how fast the variable (like $E$ or $M$) estimates saturates to their equilibrium average and how discarding initial $10^5$ MCS data can be enough for calculating the thermal averages.
\begin{figure}[ht]
  \includegraphics[width=\linewidth,keepaspectratio]{therm.eps}\\
  \vskip .2 in
  \includegraphics[width=.75\linewidth,keepaspectratio]{autocor.eps} 
         \caption{{(Color online) (a) Energy and (b) magnetization per site measured as a function of MCS for the parameters shown. (c) gives the autocorrelation function with Montte Carlo time.}}
     	 \label{therm}
     \end{figure}}
        
{One can also investigate the independence of an measurement after thermalization via computing the autocorrelation function $C(\tau)$. This accounts for the statistical memory of an observable along the Monte Carlo history, i.e., how strongly a configuration at Monte Carlo time $t$ is correlated with another configuration after an interval $\tau$. The autocorrelation function is expressed as,
	\[
	C(\tau) = \frac{\langle E(t)E(t+\tau)\rangle - \langle E\rangle^2}{\langle E^2\rangle - \langle E\rangle^2}.
	\]
	Where $E(t)$ is the energy measured at Monte Carlo step $t$, $\tau$ is the interval time and $E(t+\tau)$ is the energy at time $(t+\tau)$.
        For a normalized variable such as the energy, $C(\tau)=1$ at $\tau=0$, and it decays toward zero as the separation between MC steps increases ($\tau$ increases), indicating loss of memory and the approach to statistically independent samples. In the present case, the autocorrelation drops sharply at small $\tau$, falling below $0.1$ within a short interval and becoming essentially indistinguishable from zero at larger $\tau$, with only small fluctuations around the baseline. This rapid decay shows that the system decorrelates efficiently and that the Markov chain does not retain long-range memory over Monte Carlo time. The near-zero plateau at large $\tau$ further suggests that the chosen sampling interval is sufficient to produce approximately independent measurements after equilibration. Autocorrelation analysis is therefore necessary because it provides a quantitative estimate of the correlation time, helps identify how many Monte Carlo sweeps should be discarded as thermalization, and determines the spacing needed between successive measurements to obtain reliable error estimates.}

   \end{appendix}
   
   \section*{Acknowledgements}
   The authors thank Joao Antonio Plascak, G. Albuquerque Silva, Arghya Taraphder and Tushar Kanti Bose for fruitful discussions. Besides, RB expresses gratitude to Dwipesh Majumder for guidance.  Furthermore, financial assistance from IIEST, Shibpur, India and ANRF (DST-SERB scheme no. CRG/2022/002781), Government of India are acknowledged by RB and SK respectively.


\end{document}